\documentclass[fleqn,twoside]{article}
\usepackage{espcrc2}

\usepackage{graphicx}

\newcommand{\AmS}{{\protect\the\textfont2
  A\kern-.1667em\lower.5ex\hbox{M}\kern-.125emS}}

\title{Interplay between single-particle and two-particle tunneling in
  normal metal-$d$-wave superconductor junctions probed by shot noise}

\author{T. L\"ofwander,\thanks{E-mail address: tfstomas@fy.chalmers.se}
  V. S. Shumeiko, and G. Wendin\\
  Department of Microelectronics and Nanoscience, Chalmers University
  of Technology and G\"oteborg University, S-412 96 G\"oteborg,
  Sweden}

\begin{document}

\begin{abstract}
  We discuss how life-time broadening of quasiparticle states
  influences single- and two-particle current transport through
  zero-energy states at normal metal/$d$-wave superconductor
  junctions. We distinguish between intrinsic broadening (imaginary
  part $\eta$ of the energy), which couples the bound states with the
  superconducting reservoir, and broadening due to leakage through the
  junction barrier, which couples the bound states with the normal
  metal reservoir. We show that shot noise is highly sensitive to the
  mechanism of broadening, while the conductance is not. In the limit
  of small but finite intrinsic broadening, compared to the junction
  transparency $D$, $\eta/\Delta_0\ll D$, the low-voltage shot noise
  at zero frequency and zero temperature becomes proportional to the
  magnitude $\eta$ of intrinsic broadening
  ($\Delta_0$ is the maximum $d$-wave gap).
\end{abstract}

\maketitle

\section{Introduction}
Tunneling experiments have for a long time been an important probe of
superconducting properties\cite{Giaever}. According to early
theoretical analyses in terms of the tunneling Hamiltonian
model\cite{Bardeen}, the conductance of a normal metal-$s$-wave
superconductor tunnel junction is proportional to the superconductor
density of states. Within this model only single-particle tunneling is
taken into account. For more transparent interfaces, Blonder, Tinkham
and Klapwijk\cite{BTK} (BTK) presented a way to calculate the
conductance within Bogoliubov-de Gennes quantum mechanics. Their
approach is valid for arbitrary transparency $D$ of the interface, but
assumes that the junction system has no impurities and that inelastic
scattering is negligible.  BTK showed that for intermediate-to-high
transparency, so-called Andreev reflection\cite{Andreev} plays an
important role in determining the conductance. Andreev reflection is
the process where an electron with an energy near the Fermi surface,
incident on a superconductor from a normal metal, is converted into a
hole with essentially unchanged momentum, but with opposite group
velocity. The reversed process is also possible. In the Andreev
reflection process a Cooper pair is formed on the superconductor side
and a charge $2e$ is transferred over the NS interface, which
corresponds to a two-particle tunneling event\cite{SchWil}. In the
low-transparency limit, Andreev reflection is suppressed since it is
of second order in the transparency $D$, while ordinary single
particle tunneling is of first order in $D$. Therefore, in the tunnel
limit, $D \ll 1$, the BTK result reduces to the tunnel model result.

In contrast to the clear situation for $s$-wave junctions described
above, the interpretation of the conductance of $d$-wave NIS
junctions\cite{Geerk,Cov,Wei} where interface resonant states are
present is more complicated. Surface/interface states with zero energy
are necessarily formed as a consequence of the $d$-wave symmetry of
the order parameter, if the interface is oriented so that the order
parameter changes its sign along quasiparticle trajectories involving
normal reflection at the interface\cite{Hu}. For a recent review of
the implications of these surface/interface zero-energy states (ZES),
see Ref.~\cite{review}. It is important that two physically distinct
situations can be realized depending on the relation between (i)
intrinsic broadening due to e.g. surface roughness, impurities, or
phonons, which connects the interface states with the bulk
superconductor, and (ii) broadening due to leakage to the normal
reservoir through the barrier. When intrinsic broadening is
negligible, i.e. when the interface states are decoupled from the
superconducting reservoir, the interface states can participate in
current transport only via the Andreev reflection process. In fact,
the Andreev current is resonantly enhanced by the ZES, and is of first
order in the transparency D. On the other hand, when the intrinsic
broadening is dominating, the interface states will instead assist
single particle tunneling. Thus, for a $d$-wave superconductor with
ZES, a zero-bias conductance peak (ZBCP) is present in both limits,
but the type of tunneling responsible for it is very different. Even
if both types of life times are large, it is the quotient
\begin{equation}\label{def_q}
q = \frac{\tau_b}{\tau_r} \propto \frac{\eta}{D\Delta_0},
\end{equation}
which will determine what type of tunneling is dominating: single
particle tunneling if $q\gg 1$ and two-particle tunneling if $q\ll
1$\cite{review,wendin}. The intrinsic relaxation time
$\tau_r\propto\hbar/\eta$ is set by the damping introduced by an
imaginary part $\eta$ of the quasiparticle energy, while the life time
$\tau_b\propto\hbar/D\Delta_0$ due to leakage through the barrier is
determined by the transparency $D$ of the barrier ($\Delta_0$ is the
maximum $d$-wave gap).

In the present paper we will discuss the interplay between
single-particle and two-particle tunneling through the ZES, with
emphasis on the limit of small but finite damping. We will show that
it is possible to discriminate between the two types of tunneling in a
measurement of low-voltage shot noise at zero temperature and zero
frequency. Shot noise was recently calculated for $d$-wave NIS
junctions in Refs.~\cite{ZhuTing,Tanaka_noise}, under the implicit
assumption of negligible intrinsic broadening ($q = 0$). It was found
that the zero-voltage differential shot noise level is zero,
$(\partial S/\partial V)(V=0)=0$, because the effective Andreev
reflection probability is enhanced to unity near zero energy by the
ZES resonance. Below we will study the effect of intrinsic broadening,
and show that the noiseless character of the ZES is quickly lost when
$\eta$ increases from zero. This happens because the effective Andreev
reflection probability is reduced from unity with increasing
broadening; single-particle tunneling via the ZES then becomes
possible on the expense of two-particle tunneling, and fluctuations
between the two channels is introduced. When broadening is small, the
low-voltage shot noise will be proportional to $\eta$, which makes it
possible to probe the magnitude of intrinsic broadening by a
measurement of noise.

\section{Calculation of current and shot noise}
In our calculation we use the coherent BTK scattering approach for a
specular $d$-wave NIS junction, modified to include intrinsic
broadening on a phenomenological level by an imaginary part of the
energy. We assume that the normal metal is at $x<0$, while the
superconductor is at $x>0$. We let a $d$-wave gap node point towards
the interface, so that the spectral weight of the ZES is large. It is
known that for this orientation the gap is substantially suppressed
near the interface\cite{Nagato}. However, in order to make clear our
points on the presence of an interplay between single-particle and
two-particle tunneling as a function of $q$, and that shot noise will
be sensitive to the nature of tunneling, it is sufficient to assume a
step function form of the gap function
$\Delta(\theta,x)=\Theta(x)\Delta_0\sin 2\theta$, where $\theta$ is
the angle of quasiparticle propagation relative to the interface
normal. The exact form of the specular barrier is not important and we
assume a $\delta$-function potential characterized by a transmission
probability $D(\theta)=\cos^2\theta/(\cos^2\theta+Z^2)$, where the
dimensionless quantity $Z$ is a measure of the strength of the
potential.

\begin{figure}[t]
\includegraphics[width=6cm]{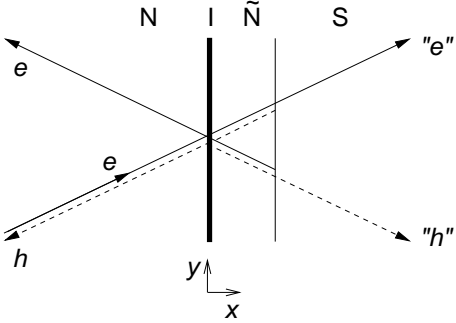}
\caption{Quasiclassical paths illustrating the structure of the
scattering state in real space (the $xy$-plane) due to an incoming
electron from the normal metal side (N). In between the barrier (I)
and the superconductor (S), we introduce a small normal region
($\tilde{\mbox{N}}$). Normal scattering takes place at the insulator
and Andreev reflection takes place at the $\tilde{\mbox{N}}\mbox{S}$
interface.}\label{xyfig}
\end{figure}

\begin{figure}[t]
\includegraphics[width=6cm]{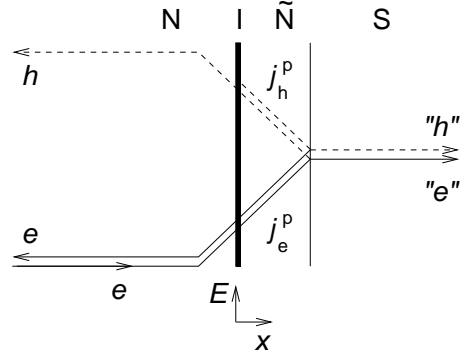}
\caption{Structure of the scattering state in energy space (the
$xE$-plane).}\label{xEfig}
\end{figure}

In the original BTK theory, effective normal reflection $R_N$ and
Andreev reflection $R_A$ probabilities are calculated and then the
charge current spectral density is expressed through them. We will
follow a slightly different route\cite{BSW}, and express the charge
current density in terms of probability current densities flowing
along the energy axis. This viewpoint makes it possible to rigorously
divide the total charge current $I(V)$ into single-particle $I_1(V)$
and two-particle $I_2(V)$ currents. When an electron at energy $E$ is
injected from the normal reservoir into the junction, it is
accelerated by the voltage drop $eV$ over the barrier. At the
superconductor it is Andreev reflected into a hole which, having the
opposite charge, is also accelerated by $eV$ when it tunnels over the
barrier back to the normal metal. Thus, the Andreev reflected hole is
at the energy $E+2eV$. Each particle carries a certain probability
current, proportional to the absolute value squared of the
wavefunction. Thus, it is natural to change focus from the charge
current flow through the junction (along the $x$-axis), to the
probability current flow along the energy axis ($E$-axis), see
Figs.~\ref{xyfig}-\ref{xEfig}. In the figures, we have introduced a
small auxiliary normal region ($\tilde{\mbox{N}}$) in between the
barrier (the insulator, I) and the superconductor (S). The final
results will not be affected by the presence of $\tilde{\mbox{N}}$ and
in the end we let the thickness of this region go to zero. On each of
the electron and hole legs in $\tilde{\mbox{N}}$ (see
Fig.~\ref{xEfig}), we can define the total probability current
densities $j^p_e$ and $j^p_h$, respectively, including the effects of
normal backscattering at the insulator. For energies outside the gap,
the Andreev reflection probability is less than unity, which leads to
a leakage $j^p_e-j^p_h$ of probability current during Andreev
reflection. The total leakage is the single-particle current. The rest
of the probability current (the part surviving Andreev reflection)
continues out into the normal metal as a hole current, and contributes
to the two-particle current. For subgap energies, the Andreev
reflection probability is unity; probability current is then conserved
during Andreev reflection, the leakage $j^p_e-j^p_h$ vanishes, and
single particle tunneling is quenched.

In the zero-temperature limit, the current then takes the
form\cite{BSW}
\begin{eqnarray}
eR_nI(V) &=& eR_n\left[I_1(V) + I_2(V)\right] \nonumber\\
&=& \frac{1}{2\left<D\right>}\int_{-\pi/2}^{\pi/2} d\theta\cos\theta\left[
I_1(V,\theta) + I_2(V,\theta)\right], \nonumber\\
I_1(V,\theta) &=& \int_{-eV}^0 dE\,
\frac{D(1-|a|^2)(1+R|\bar a|^2)}{|1-Ra\bar a|^2},\label{IV_eq}\\
I_2(V,\theta) &=& 2\,\int_{-eV}^0 dE\,
\frac{D^2|a|^2}{|1-Ra\bar a|^2},\nonumber
\end{eqnarray}
where $R_n=\pi h/e^2k_F L_y\left<D\right>$ is the normal state
junction resistance, $L_y$ is the junction width, $k_F$ is the Fermi
wave vector, $\left<D\right>=\int d\theta \cos\theta D(\theta)/2$, and
$a=a(\theta,E)$ is the Andreev reflection amplitude. The amplitude
$\bar a$ is calculated at the angle $\pi-\theta$, the angle of
propagation after normal reflection at the barrier. The differential
conductance $G(V)$ is obtained by differentiation with respect to $V$.

\begin{figure}[t]
\includegraphics[width=7cm]{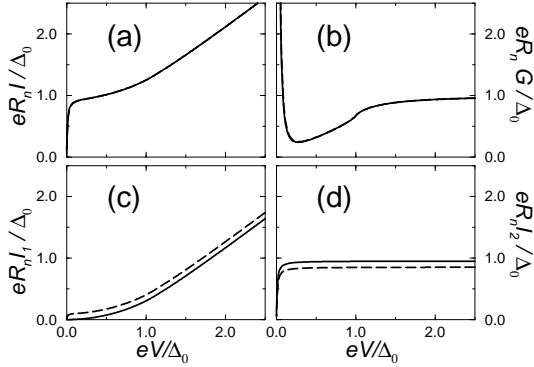}
\caption{(a) The total current, (b) the differential conductance, (c) the
  single particle current, and (d) the two-particle current, for a
  $d$-wave NIS junction oriented so that a gap node is pointing
  towards the interface. The solid lines are calculated for $\eta=0$,
  while the dashed lines are calculated for $\eta=0.001\Delta_0$. The
  zero-energy states resonantly enhance the two-particle current,
  which is of first order in the transparency.  Increasing broadening
  enhances the single-particle current on expense of the two-particle
  current, but leaves the total current unaffected. The transparency
  of the junction is $\left<D\right>=0.026$, and the temperature is
  zero.}\label{currentfig}
\end{figure}

For the calculation of shot noise we follow the technique in
Ref.~\cite{Buttiker}, see also \cite{ZhuTing}. The formula for the
noise, here limited to zero temperature and zero frequency, is
\begin{eqnarray}\label{NISnoise}
R_nS(V) &=& \frac{1}{\left<D\right>}
\int_{-\pi/2}^{\pi/2} d\theta \cos\theta
\int_{-eV}^0 dE\, S(\theta,E),\nonumber\\
S(\theta,E) &=& R_N(1-R_N) + R_A(1-R_A)\\
&& + 2R_NR_A,\nonumber
\end{eqnarray}
where $R_N=R|1-a\bar a|^2/(|1-Ra\bar a|^2$, and
$R_A=D^2|a|^2/|1-Ra\bar a|^2$. In the subgap region, for zero
temperature and negligible broadening, the only surviving source of
noise is a term involving scattering states separated in energy by
$2eV$: the first scattering state consists of an electron incoming
from the normal metal reservoir at the energy $E$, Andreev reflected
as a hole which is ejected back into the normal metal at the energy
$E+2eV$; the second scattering state consists of a hole incoming at
$E+2eV$, Andreev reflected as an electron emerging at $E$.

We plot the total current $I(V)$, the conductance $G(V)$, the
single-particle current $I_1(V)$, and two-particle current $I_2(V)$ in
Fig.~\ref{currentfig}(a)-(d). From Fig.~\ref{currentfig} it is clear that in
the absence of broadening, the solid lines ($q=0$), the ZBCP is due to
the two-particle current only. One can show that the two-particle
current is of first order in the transparency, $I_2\propto D$, because
of the ZES resonance. The width of the ZES resonance is
$\Gamma(\theta)=D(\theta)|\Delta(\theta)|/
\left[2\sqrt{R(\theta)}\right]$
for each quasiparticle trajectory angle $\theta$, see also e.g.
Ref.~\cite{review,Tanaka_rev}.

To study the effects of intrinsic broadening, the quasiclassical
Green's function technique\cite{Eliashberg} can be used. Recently, a
highly useful parameterization was introduced by Nagato, Nagai, and
Hara\cite{Nagato2} for the equilibrium case, and by
Eschrig\cite{Eschrig} for the non-equilibrium case, which can be used
to separate electron and hole parts of the Green's functions.
Probability currents flowing along the energy axis are then easily
found, and one can divide the total charge current into single- and
two-particle currents, in a one-to-one mapping to the approach
outlined above. Eventually, the result is the same expression as in
Eq.~(\ref{IV_eq}), but with the Andreev reflection amplitudes
substituted by generalized Andreev reflection amplitudes containing
the effects of damping in the form of an imaginary part of the
energy\cite{Eschrig,Poenicke}. When broadening increases, the
two-particle current is suppressed and the single-particle current is
enhanced.  However, the total current and the conductance are not
particularly affected, as shown in Fig.~\ref{currentfig}(a)-(d) by the
dashed lines. In fact, in the limit of large broadening, $q \gg 1$,
the ZBCP is solely due to single-particle tunneling. In this case, it
is convenient to instead of the above approach apply the tunnel
formula for the conductance and calculate the local density of states
at the surface, as in e.g.  Ref.~\cite{MatShi}. In this limit the
width of the ZBCP is directly proportional to the magnitude of the
broadening $\eta$.

\begin{figure}[t]
\includegraphics[width=7cm]{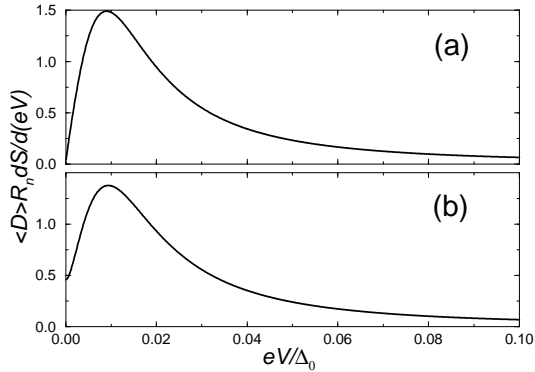}
\caption{Differential shot noise $\partial S/\partial V$ at zero
  temperature and zero frequency, calculated for the same junction as
  in Fig.~\ref{currentfig}. In (a) for $\eta=0$, and in (b) for
  $\eta=0.001\Delta_0$. The noiseless character of the zero-energy
  states is quickly lost when broadening is
  introduced.}\label{noisefig}
\end{figure}

In Fig.~\ref{noisefig} we plot the differential shot noise, $\partial
S/\partial V$. The ZES resonance does not produce any noise $(\partial
S/\partial V)(V=0)=0$, see Fig.~\ref{noisefig}(a), as found in
Refs.~\cite{ZhuTing,Tanaka_noise}.  However, when a small amount of
intrinsic broadening is introduced, the noiseless character of the ZES
is lost, see Fig.~\ref{noisefig}(b). This behavior can be understood
in the following qualitative way. When there is no broadening, the ZES
resonantly enhances the Andreev current. The effective Andreev
reflection probability is unity, despite the fact that $D \ll 1$, c.f.
Eq.~(\ref{IV_eq}) for $E=0$ and Fig.~\ref{currentfig}(d). The result
of zero noise follows directly from Eq.~(\ref{NISnoise}), since the
relation $R_N+R_A=1$ holds in the subgap region (probability current
conservation). When broadening increases from zero, the effective
Andreev reflection probability is reduced from unity, and the noise
becomes finite. In the limit of small intrinsic broadening, $q\ll 1$,
the noise at zero-voltage is proportional to the magnitude of
broadening, $(\partial S/\partial V)(\theta,V=0)\propto
\eta/\Gamma(\theta)$, where $\Gamma(\theta)$ is the width of the ZES
in the absence of damping.

\section{Summary}
We have discussed the interplay between single-particle and
two-particle tunneling through zero-energy states (ZES) in $d$-wave
NIS junctions. For small intrinsic broadening of the ZES, compared to
the broadening due to leakage over the barrier to the normal metal
reservoir, $q \ll 1$, current is transported via two-particle
tunneling. On the other hand, for large damping, $q \gg 1$, only
single particle tunneling is present. We have shown that shot noise is
highly sensitive to the type of tunneling, although the conductance is
not. The noise-less character of the ZES found in
Ref.~\cite{ZhuTing,Tanaka_noise} is quickly lost when damping is
introduced. For small but finite intrinsic broadening, $q\ll 1$, the
low-voltage shot noise is directly proportional to the magnitude of
broadening.

\section*{Acknowledgments}

It is a pleasure to thank Mikael Fogelstr\"om for valuable
discussions.  This work has been supported by grants from NEDO
(Japan), NFR (Sweden), and SSF (Sweden-Japan QNANO project).

\end{document}